\documentclass[a4paper,fleqn,usenatbib]{mnras}
\pdfoutput=1

\usepackage[pdftex]{graphicx,color}
\usepackage[T1]{fontenc}
\usepackage[utf8]{inputenc}
\usepackage{lmodern}
\usepackage{float}
\usepackage{caption}

\definecolor{urlblue}{rgb}{0,0,0.9}
\definecolor{linkgreen}{rgb}{0,0.45,0}
\definecolor{linkorange}{rgb}{0.7,0.1,0.0}

\usepackage{amsmath, amssymb}
\usepackage{cuted}
\usepackage[english]{babel}
\usepackage{enumerate}
\usepackage[normalem]{ulem}
\usepackage{enumitem}
\usepackage{multirow}
\usepackage{booktabs}
\usepackage{textcase}
\usepackage{makecell}

\setlist[enumerate]{wide=0pt, widest=99,leftmargin=\parindent, labelsep=* } 

\definecolor{valecol}{rgb}{0,0.5, 1.}

\definecolor{rancol}{rgb}{1,0., 0.}

\def\d{{\rm d}}

\graphicspath{{./figs/}}

\AtBeginDocument{\hypersetup{
linkcolor=linkgreen,
citecolor=linkorange,
urlcolor=urlblue}}

\title[BAO in thin redshift shells]{Baryon acoustic oscillations in thin redshift shells from BOSS DR12 and eBOSS DR16 galaxies}

\author[Menote \& Marra]{
Ranier Menote$^{1}$ and Valerio Marra,$^{2,3,4}$
\\
$^{1}$PPGCosmo, Universidade Federal do Espírito Santo, 29075-910, Vitória, ES, Brazil\\
$^{2}$Núcleo de Astrofísica e Cosmologia \& Departamento de Física, Universidade Federal do Espírito Santo, 29075-910, Vitória, ES, Brazil\\
$^{3}$INAF -- Osservatorio Astronomico di Trieste, via Tiepolo 11, 34131, Trieste, Italy\\
$^{4}$IFPU -- Institute for Fundamental Physics of the Universe, via Beirut 2, 34151, Trieste, Italy\\
}

\begin{document}

\label{firstpage}
\pagerange{\pageref{firstpage}--\pageref{lastpage}}

\maketitle

\begin{abstract}
In an age of large astronomical datesets and severe cosmological tensions, the case for model independent analyses is compelling.
We present a set of 14 baryon acoustic oscillations measurements in thin redshift shells with $3\%$ precision that were obtained by analyzing BOSS DR12 and eBOSS DR16 galaxies in the redshift range $0.32<z<0.66$.
Thanks to the use of thin shells, the analysis is carried out using just redshifts and angles so that the fiducial model is only introduced when considering the mock catalogs, necessary for the covariance matrix estimation and the pipeline validation.
We compare our measurements, with and without supernova data, to the corresponding constraints from Planck 2018, finding good compatibility.
A Monte Python module for this likelihood is available at \href{https://github.com/ranier137/angularBAO}{github.com/ranier137/angularBAO}.
\end{abstract}

\begin{keywords}
large-scale structure of Universe -- cosmology: observations -- cosmological parameters -- cosmology: theory
\end{keywords}

\section{Introduction}\label{sec:intro}

Observational cosmology is being characterized by ever larger datasets, but also ever worse discrepancies in the predictions of its standard model \citep{Perivolaropoulos:2021jda}, as remarked by the $5\sigma$ tension between late and early determinations of the Hubble constant~\citep{Riess:2021jrx}.
This calls for model independent methods for analyzing observational data.

Model independent observational constraints are generally weaker than the ones that adopt assumptions to reduce the degrees of freedom of a problem. Sufficiently large dataset are thus necessary. Common hypotheses regard the metric and content of the universe. Regarding the former, one usually assumes large-scale homogeneity and isotropy, and so the FLRW metric. Regarding the latter, raw data are usually analyzed withing the framework of the $\Lambda$CDM model, often fixing its parameters to the fiducial standard model values.
This means that model independent methods allow one to test deviations from the FLRW metric \citep{Camarena:2021mjr} and the $\Lambda$CDM paradigm \citep{DiValentino:2021izs} in a more consistent way.
In particular, model independent analyses could be useful to understand the causes of the standard model tensions, exactly because it is easier, with less assumptions,  to narrow down the problem.

Here, we present a set of 14 baryon acoustic oscillations (BAO) measurements in thin redshift shells with $3\%$ precision that were obtained by analyzing BOSS DR12 and eBOSS DR16 galaxies in the redshift range $0.32<z<0.66$.
Thanks to the use of thin shells \citep{Sanchez:2010zg}, the analysis is carried out using just redshifts and angles so that the fiducial model is only introduced when considering the mock catalogs, necessary for the covariance matrix estimation and the pipeline validation.%
\footnote{See \citet{Marra:2018zzo} for the estimation of the radial BAO signal using only angles and redshifts.}

We compare these angular BAO measurements, with and without supernova data, to the corresponding constraints from Planck 2018, finding good compatibility.
BAO measurements in thin redshift shells were also obtained by the Observatório Nacional Group \citep[ON Group, ][]{Carvalho:2017tuu,Carvalho:2015ica,Alcaniz:2016ryy}, which, however, found a tension with respect to CMB results.
We also combine the angular BAO and supernova data with the SH0ES prior on $M$. We find that it causes a strong $4.5\sigma$ tension in the $r_d$-$H_0$ plane with respect to Planck 2018: it is the $H_0$ crisis.
A Monte Python module for this likelihood is available at \href{https://github.com/ranier137/angularBAO}{github.com/ranier137/angularBAO}.

This paper is organized as follows. In Section~\ref{sec:data} we describe the data, and in Section~\ref{sec:method} our methodology. Our set of 14 BAO measurements in thin redshift shells is obtained in Section~\ref{sec:bao} and cosmological inference is carried out in Section~\ref{sec:cosmo}. We conclude in Section~\ref{sec:conclusions}.

\section{Data}\label{sec:data}

\begin{figure}
\centering 
\includegraphics[trim={0 0 0 0}, clip, width=1 \columnwidth]{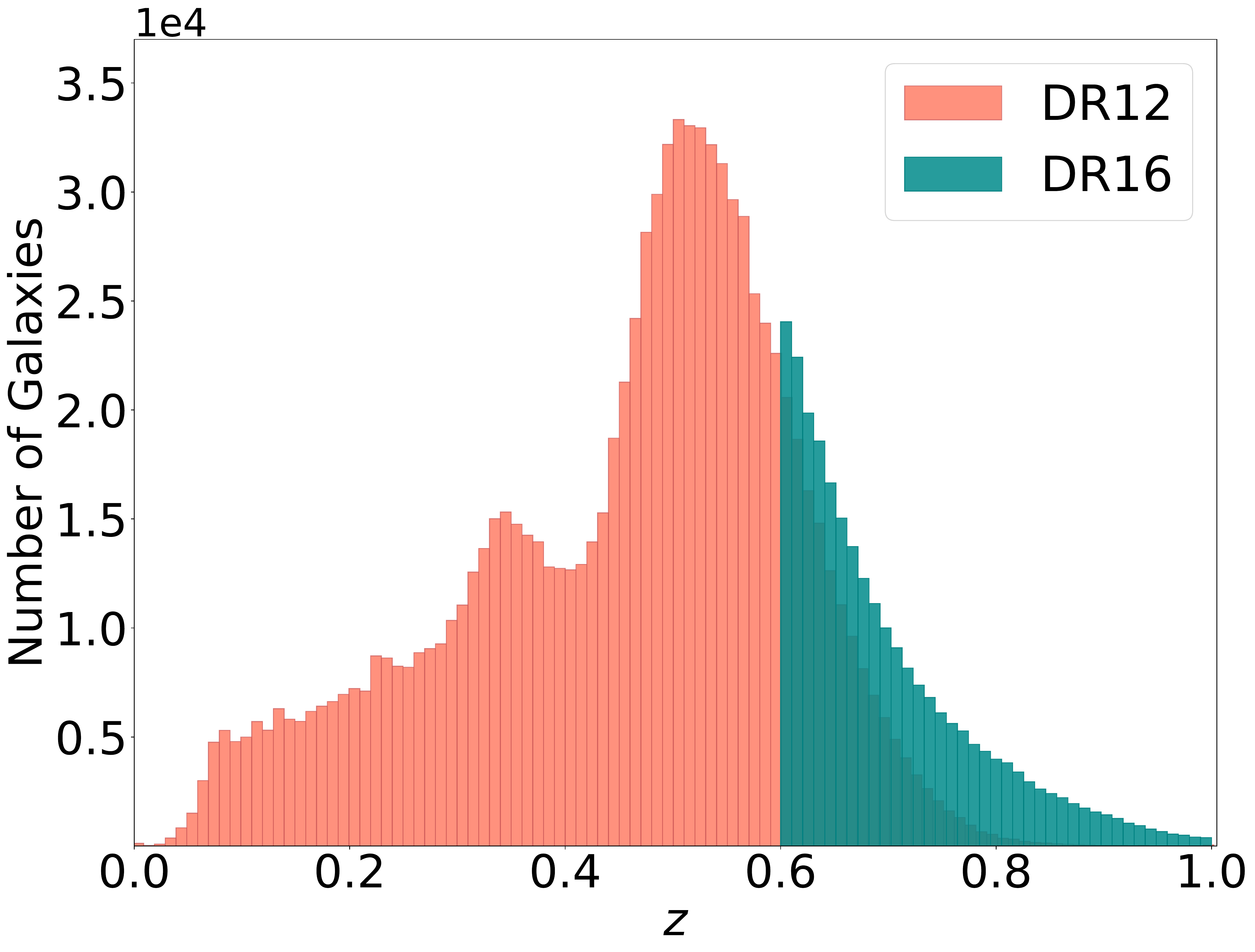}
\caption{Redshift distributions of the luminous red galaxies from BOSS DR12 and eBOSS DR16.
We considered the redshift range $0.32<z<0.66$.}
\label{fig:boss}
\end{figure}

The \textit{Sloan Digital Sky Survey} (SDSS) is an international scientific collaboration that has created the most detailed three-dimensional maps of the Universe. This project was divided into 4 phases, SDSS-I (2000-2005), SDSS-II (2005-2008), SDSS-III (2008-2014), and SDSS-IV (2014-2020).  This work deals with the last \textit{Data Release} (DR) 12 of the \textit{Baryon Oscillation Spectroscopic Survey} \citep[BOSS,][]{2013AJ....145...10D} and the DR16 of the \textit{extended Baryon Oscillation Spectroscopic Survey}  \citep[eBOSS,][]{2016AJ....151...44D}, subsets of SDSS-III and SDSS-IV, respectively.
Fig.~\ref{fig:boss} shows their redshift distributions.
We considered the catalogs of luminous red galaxies (LRGs) that cover the northen sky.
We do not consider the southern sky because the smaller number of galaxies does not allow for a robust determination of the BAO signal in thin redshift slices.
DR12-North offers a catalog with about 953 thousand LRGs with redshift up to $z = 0.8$, while DR16-North contains about 256 thousand LRGs within the redshift range $0.6 < z < 1.0$.

In order to estimate the angular correlation function we use random catalogs from BOSS and eBOSS about 50 and 20 times larger than the corresponding real ones, respectively. In order to estimate the covariance matrix and the robustness of the method we use 1000 mock catalogs from BOSS \citep[\texttt{Patchy-Mocks-DR12NGC-COMPSAM V6C}, ][]{2016MNRAS.456.4156K,2016MNRAS.460.1173R} and eBOSS \citep[\texttt{EZmock\_eBOSS\_LRGpCMASS\_NGC}, ][]{Zhao:2020bib}.

\section{Method}\label{sec:method}

As said earlier, we wish to estimate the projected correlation function $w(\theta)$ in thin redshift shells, which is related to the 3D correlation function $\xi(s)$ via:
\begin{align}
w(\theta, z) & \xrightarrow{\text{thin-shell limit}}\xi \big( s(\theta,z), z \big ) \,, \label{w-theo} \\
s(\theta, z) &= 2 \, (1+z) d_A(z) \sin \frac{\theta}{2} \,,
\end{align}
so that the BAO peak is at the scale:
\begin{align}\label{cosmo_BAO}
\theta_{\rm BAO}(z) = \frac{r_d}{(1+z) d_A(z)} \,,
\end{align}
where $r_d$ is the sound horizon at the drag epoch and $d_A$ is the angular diameter distance. Our goal  is to estimate $\theta_{\rm BAO}$.

Here, we focus on the BAO peak of the angular correlation function, that is, we consider only the background expansion. Alternatively, one could use the correlation function multipoles to constrain also the growth of structures, see \citet{Wang:2017wia}.
Furthermore, in order to avoid converting redshifts and angles via a fiducial model, one could constrain the growth of the large-scale structure via the angular power spectrum as proposed by \citet{Tanidis:2021uxp,Camera:2018jys}.


For later use, considering the late-time flat $\Lambda$CDM model, we list the relevant background quantities:
\begin{align}
r(z) & = \int_0^z \frac{c \, \d \bar z}{H(\bar z)}  \,, \\
{H^2(z) \over H_{0}^{2} } &= \Omega_{m}   (1+z)^{3}+ 1 - \Omega_{m}  \,, \\
d_A(z) &= (1+z)^{-1} r(z) \,, \\
d_L(z) &= (1+z) \, r(z) \,, \\
m (z) & = 5\log_{10}\left[\frac{d_L(z)}{10 \text{pc}} \right]  + M \,, \\
\mu(z)&= m(z) - M \,,
\end{align}
where we defined the comoving distance, the Hubble rate, the angular diameter distance, the luminosity distance, the apparent magnitude and the distance modulus, respectively. The parameter $M$ is the absolute magnitude of the source.

\subsection{Angular Correlation Function}

The $i$-th galaxy of the catalog has redshift $z_i$ and angular position $\beta_i = (\alpha_i, \delta_i)$, where $\alpha_i$ and $\delta_i$ are its right ascension and declination, respectively.
Considering two galaxies within the redshift bin $[z -\frac{\delta z}{2}, z +\frac{\delta z}{2}]$, their angular separation $\gamma$ is given by:
\begin{align}
\gamma (\beta_i, \beta_j) = \arccos \Big(&\sin(\delta_i)\sin(\delta_j) +\nonumber\\
& \cos(\delta_i)\cos(\delta_j) \cos(\alpha_i-\alpha_j) \Big) \,,
\label{angularseparation}
\end{align}
with which one can estimate the \textit{two-point angular correlation function} (ACF) by counting pairs in various angular bins.

The catalogs give statistical weights associated to each galaxy in order to correct for possible observational biases. Galaxies are then counted according to the total weight \citep{Anderson:2012sa}:
\begin{eqnarray}
 w = w_{_{\rm FKP}}w_{\rm sys}\left(w_{\rm rf} + w_{\rm cp}-1\right) \,,
\end{eqnarray}
where $w_{\rm sys}=w_{\rm star}w_{\rm seeing}$ is the total systematic weight, $w_{\rm rf}$ is the redshift failure weight, and $w_{\rm cp}$ is the close pair weight.
The FKP weights are meant to optimally weight the galaxies \citep[][]{1994ApJ...426...23F}. In the present case of thin redshift shells they have at most a 10\% variation.
Then, the total count of pairs separated by an angular scale within the interval $I_k \equiv [\theta_{k},\theta_k +\Delta \theta[$ is given by $\sum_{i,j>i}\phi_{\theta_k}(\beta_i, \beta_j)$, where the counting function is defined according to:
\begin{eqnarray}
  \phi_{\theta}(\beta_i,\beta_j) = \begin{cases}   w_iw_j &\theta \le \gamma(\beta_i,\beta_j) < \theta+\Delta \theta    \\ 0, & \text{otherwise} \end{cases} \,.
  \label{new counting function}
\end{eqnarray}
%

Considering the real catalog, the normalized count of pairs of galaxies relative to an angular separation  within the interval $I_k$ is:
\begin{eqnarray}
 DD(\theta_k) \equiv 2\frac{\sum_{i, j>i} w_i w_j}{\left(\sum_l w_l \right)^2 - \sum_l w_l^2} \text{\hspace{1mm} for \hspace{1mm}} \gamma_{ij} \in I_k \,,
\end{eqnarray}
so that $\sum_k DD(\theta_k)=1$.

In order to estimate the correlation function we adopt the optimal estimator by \citet{1993ApJ...412...64L}. Therefore, we need the normalized counts relative to the random catalog:
\begin{eqnarray}
 RR(\theta_k) \equiv 2\frac{\sum_{i, j>i} w_i w_j}{\left(\sum_l w_l \right)^2 - \sum_l w_l^2} \text{\hspace{1mm} for \hspace{1mm}} \gamma_{ij} \in I_k \,,
\end{eqnarray}
and also the cross count between the real and random catalog:
\begin{eqnarray}
   DR(\theta_k) = \frac{\sum_{i,j} w_i w_j}{\sum_{l,m} w_lw_m} \text{\hspace{1mm} for \hspace{1mm}} \gamma_{ij} \in I_k  \,,
\end{eqnarray}
where the $i$-galaxy belongs to the real catalog while the $j$-galaxy to the random one.
Finally, the angular correlation function is estimated according to:
\begin{eqnarray}
   w(\theta_k) = \frac{DD(\theta_k)-2DR(\theta_k) + RR(\theta_k)}{RR(\theta_k)} \,.
   \label{angularcorreestimator}
\end{eqnarray}

\subsection{Covariance matrix}

The data analysis requires an accurate estimation of the covariance between $w(\theta_k)$ and $w(\theta_{k'})$.
We then use Eq.~\eqref{angularcorreestimator} to compute the angular correlation function for the 1000 mock catalogs described in Section~\ref{sec:data}.
The covariance matrix is then estimated according to:
\begin{eqnarray} \label{covi}
S_{kk'} = \frac{1}{N\!\!-\!\!1}\! \sum_{m=1}^N [w_{m}(\theta_k)\!-\!\overline{w}(\theta_k)][w_{m}(\theta_{k'})\!-\!\overline{w}(\theta_{k'})] \,,
\end{eqnarray}
where $w_m(\theta_k)$ is the correlation function for the $m$-th mock catalog, $\overline{w}(\theta_k)$ is the mean, and $N=1000$ is the number of mock catalogs.
As the estimation of the covariance matrix is based on mock catalogs, it depends to some extent on the fiducial cosmology that was adopted.
This introduces a model-dependent component in the analysis.

To avoid using model-dependent mock catalogs, the covariance matrix could be estimated by splitting the complete dataset into many sub-samples.
While the galaxy density from BOSS and eBOSS is not high enough to carry out such an analysis, future datasets by DESI \citep{DESI:2016fyo}\footnote{\href{https://www.desi.lbl.gov}{www.desi.lbl.gov}} and Euclid \citep{Amendola:2016saw}\footnote{\href{https://www.euclid-ec.org}{www.euclid-ec.org}} could be suitable.
Alternatively, one could adopt the method developed by \citet{Angulo:2009rc,Contreras:2020kbv} in order to change the cosmology of a halo sample and study the impact of the choice of a fiducial model on cosmological inference.

\subsection{Phenomenological model}

Because of gravitational interaction, it is expected that the correlation of galaxy positions is larger at smaller scales, then decreasing, like a power law, towards larger scales, reaching eventually homogeneity ($w\approx 0$). However, we should see, at intermediate scales, a sudden increase in the correlation, marking the BAO feature.
Following \citet{Sanchez:2010zg}, we  use the following phenomenological model in order to constrain the BAO peak position directly from the data:
\begin{equation}
    w(\theta) = A + B\theta^{\nu} + Ce^{-\frac{(\theta-\theta_{\rm fit})^2}{2\sigma^2}} \,.
    \label{phenom_model}
\end{equation}
We consider angular bins with $\Delta \theta = 0.3^{\circ}$, which are fine enough so as to resolve the BAO feature, but wide enough in order to reduce shot noise. We compute the correlation functions in the range $1.5^{\circ} < \theta < 7^{\circ}$.

In the above parametrization, the only physically interesting parameters are the BAO position $\theta_{\rm fit}$ and the BAO signal strength $C$. The other parameters are nuisance parameters, to be marginalized over, that are necessary to reach a satisfactory fit to the data and whose correlations with $\theta_{\rm fit}$ will affect the final constraint on the BAO signal.
We tested more complicated templates for the overall shape of the correlation function. We examined the double exponential template of \citet{Sanchez:2012eh}, one with a double Gaussian and also one with a coupling between the exponential and Gaussian terms. We found that data does not justify them, as these templates do not provide better results relative to their increased  number of parameters.

\subsection{Detection of the BAO signal and priors}

The chi-squared function is defined as
\begin{eqnarray} \label{chi2-mes-bao}
\chi^2 = \sum_{ij}
\left[w_i-w(\theta_i)\right] S^{-1}_{ij}\left[w_j-w(\theta_j)\right]   \,,
\end{eqnarray}
where $w_i$ are the measurements, $w(\theta_i)$ is the model of Eq.~\eqref{phenom_model} and $S_{ij}$ is the covariance matrix of Eq.~\eqref{covi}. 

We adopt wide-enough flat priors on the parameters $A$, $B$ and $\nu$ that describe the correlation function without BAO feature.
Regarding the parameters describing the BAO peak, we adopt a flat prior on $C$ and $\theta_{\rm fit}$, but an informative flat prior on $\sigma$: $0.15^\circ< \sigma< 1^\circ$.
The lower bound is justified by the binning width that we adopted as the narrowest Gaussian peak we can detect satisfies $4\sigma \sim 2\Delta \theta$, that is, the 4$\sigma$ extension of the Gaussian would cover two angular bins (three points $\theta_k$).
The upper limit comes from theoretical expectations, fitting the template of Eq.~\eqref{phenom_model} to Eq.~\eqref{w-theo}.
Ideally, one should use a left-bounded prior for $\sigma$ but the number of galaxies is not high enough to permit such an analysis.
The angular scale that maximizes the marginalized posterior on $\theta_{\rm fit}$ is denoted by $\theta_{\rm mc}$.

As said earlier, $C$ parametrizes the BAO signal strength and it is expected to be positive. Shot noise, however, could produce false signals of positive or negative intensity. Therefore, we use the posterior on $C$ in order to assess the detection of the BAO peak.
More precisely, we define the detection strength according to: 
\begin{align}\label{S-BAO}
S_{\rm BAO} = \int_0^\infty f(C) \, \d C \,,
\end{align}
where $f(C)$ is the posterior on $C$. We consider only measurements that satisfy $S_{\rm BAO} > 0.95$. 
In other words, we consider only measurements for which the null hypothesis that the BAO feature is absent is excluded at more than 95\% confidence level.


\subsection{Biases in the estimation of the BAO peak}
\label{sec:biases}

We now discuss the possible sources of bias that affect the determination of the angular scale of the BAO peak.
The major sources of bias are due to projection effects, modeling and parametrization assumptions, and redshift space distortions.
We will discuss their magnitude and how we took them into account in the next Sections.

\subsubsection{Projection bias}
\label{sec:projection}

An important bias that needs to be considered is the one caused by the finite width of the redshift bins.
The number of galaxies in the BOSS and eBOSS catalogs is not sufficient in order to consider a redshift shell of negligible thickness. 
Projection effects will then displace the BAO peak to smaller angles.
\citet{Sanchez:2010zg} showed that this shift depends on the bin width, its redshift and, to a smaller degree, on the underlying cosmological model.

One could then proceed in two ways. The first approach is to deliver the $\theta_{\rm BAO}(z)$ value relative to the bin $\Delta z$ that was used. In other words, when comparing data to model, one should correct the theoretical BAO angle of Eq.~\eqref{cosmo_BAO} according to the projection effect relative to the bin $\Delta z$, redshift and cosmology that are considered. This would require estimating the correlation function from the power spectrum after applying the relevant selection function for every point of the parameter space.
The second approach is to deliver the $\theta_{\rm BAO}(z)$ value relative to $\Delta z = 0$ so that  cosmological inference can be carried out by simply considering Eq.~\eqref{cosmo_BAO}.
We adopt the second approach as it offers a much simpler likelihood.

Here, we try to estimate the projection bias directly from the data, without adopting a cosmological model. 
More precisely, for a given  redshift $z$ we estimate the correlation function for the following range of bin widths $\delta z_i$:
\begin{align}
\{ \delta z_i \} \!\!=\!\! \{0.005,0.0075, 0.01,0.0125,0.015,0.0175,0.02\} ,
\end{align}
and study how $\theta_{\rm mc}$ depends on $\delta z$. Note that adjacent values of $z$ are displaced by 0.02 in order to avoid overlap. 

As the highest value of $\delta z$ is much smaller than unity, we can expand the dependency of $\theta_{\rm mc}$ on $\delta z$ in a Taylor series up to second order.
Based on geometrical arguments,%
\footnote{Because of the finite thickness of the bin, the projected BAO scale is proportional to $\cos  \epsilon \sim 1 - \epsilon^2/2$, where $\epsilon \propto \delta z$ is the angle with respect to the tangent to the redshift shell.}
also confirmed by the results of \citet{Sanchez:2010zg}, the first derivative of this correction at $\delta z = 0$ is null so that the expansion has only two free parameters:
\begin{eqnarray}
\theta(z,\delta z) = \theta_0(z) + E(z)\delta z^2 \,.
\label{taylor_expansion}
\end{eqnarray}
In order to correct for projection, we need to determine the function $E(z)$.
We adopt a linear model:
\begin{align} \label{eq:Ez}
E(z) = E_0 + E_1 z \,,
\end{align}
but we found that the data does not justify $E_1$, which we then set to zero.

The function $\theta_0(z)$ is  meant to absorb the cosmology dependence of $\theta_{\rm BAO}(z)$. To this end we model it as a piece-wise function:
\begin{align} \label{theta-z}
\theta_{0}(z)  =
\begin{cases}
\dots & \dots \\
\theta_{0,j} & \text{ if } z=z_j \\ 
\dots & \dots
\end{cases} \,.
\end{align}
Each redshift is given its own parameter in order not to force an ansatz on the redshift dependence of $\theta_0(z)$.

We then adopt the following chi-squared function:
\begin{align} \label{chi2E}
\chi^2 (\{\theta_{0,j}\}, E_0 )&=
\sum_j  \chi_j^2 (\theta_{0,j}, E_0 )  \,, \\
\chi_j^2 (\theta_{0,j}, E_0 ) &=
\sum_{ik} 
\big [\theta_{{\rm mc},ij}- \theta_{ji} \big ] \Sigma_{j,ik}^{-1} \big [\theta_{{\rm mc},kj}- \theta_{jk} \big ] \,,
\end{align}
where $\theta_{ji} = \theta(z_j,\delta z_i)$, $i$ labels the thickness $\delta z_i$ and $j$ the redshift $z_j$.
We include only measurements for which $S_{\rm BAO}>0.95$ so that the covariance matrix $\Sigma$ ranges from $7\times 7$ to a single entry.
The covariance matrix is obtained via:
\begin{align}
\Sigma_j = \text{diag}\big(\{\sigma_{{\rm mc},ij}\}\big) \, R_{{\rm mc},j} \, \text{diag}\big(\{\sigma_{{\rm mc},ij}\}\big) \,,
\end{align}
where $\sigma_{{\rm mc},ij}$ are the uncertainties from the posterior on $\theta_{{\rm mc},ij}$ and $R_{{\rm mc},j}$ is the correlation matrix relative to the $\{ \delta z_i \}$. We estimate the correlation matrix $R_{{\rm mc},j}$ by minimizing the $\chi^2$ of Eq.~\eqref{chi2-mes-bao} for the 1000 mock catalogs, keeping only the fits with $C>0$.
Note that, for different redshift $z_j$, the $\theta_{{\rm mc},ij}$ are independent because we adopted non-overlapping redshift slices.

Summarizing, the constraints on $\theta_{\rm BAO}(z)$ relative to $\Delta z=0$ are given by the posteriors on $\theta_{0,j}$ that are obtained from the chi-squared function of Eq.~\eqref{chi2E}.



\subsubsection{Inference bias}
\label{mocks_verification}

To test the validity of the template of Eq.~\eqref{phenom_model} we fit it to the mean of the correlation functions from the 1000 mocks. In this case we use the chi-squared function of Eq.~\eqref{chi2-mes-bao} with the covariance matrix relative to the mean, that is, $S/1000$.
In order to neglect projection effects we consider the thinnest shell with $\delta z = 0.005$.
We found that across the various redshift bins this inference bias $B$ is approximately constant and of magnitude\footnote{The same result was found when considering the projection effect.}:
\begin{align}
B = \frac{\theta_{\rm mc}}{\theta_{\rm fid}} -1= (3.2 \pm 0.1)\% \,.
\end{align}
We checked that this bias can be reduced with a more complicated model, which, however, would not be justified by the actual data: the extra parameters would lead to poor constraints on most of the parameters of the model.

Considering this bias, the final measurement of the BAO peak that we report in this work is given by the expression:
\begin{eqnarray}
\theta_{\rm BAO} = \frac{\theta_0 (z)}{1+B} 
\label{final_bao_report}
\end{eqnarray}
The error on $B$ is included in the error budget.

\subsubsection{Further bias and systematic errors}
\label{bias}

As pointed out by \citet{Sanchez:2010zg}, there are other sources of bias and systematic errors that are relative to the approach adopted in this analysis.
The first one is associated to the choice of the angular interval that is used to measure the angular correlation function. Moving, for example, the starting point to smaller angles, the fitting procedure will be exposed to higher correlation and this may shift the recovered position of the BAO peak.
This parametrization bias $\delta \theta_{\rm par}$ is around  $1\%$.

Then, the nonlinear gravitational growth of structure, scale dependent and non-local bias, and redshift-space distortions also introduce a systematic bias $\delta \theta_{\rm rsd}$ in the determination of $\theta_{BAO}$, which is also around 1\% \citep{Sanchez:2010zg}.
See \citet{Prada:2014bra} for a thorough discussion of 
possible systematic shifts and damping in baryon acoustic oscillations due to the effects above.

\subsection{Total error budget}

The total uncertainty in the determination of the BAO scale includes all the sources of error that were previously discussed:
\begin{align}
\sigma_{\rm sys} & = \sqrt{\delta \theta_{\rm par}^2 +\delta \theta_{\rm rsd}^2 +\delta \theta_{B}^2} \,, \\
\sigma_{\rm BAO} &= \sqrt{\sigma_{\rm stat}^2+ \sigma_{\rm sys}^2} \,,
\label{total_errors}
\end{align}
where $\sigma_{\rm stat}$ is the uncertainty from the posterior on $\theta_0$, $\delta \theta_{\rm par}=\delta \theta_{\rm rsd}= 0.01 \, \theta_0$, and $\delta \theta_{B} \simeq 0.001 \, \theta_0$.
The full covariance matrix is given by:
\begin{align} \label{covabao}
\Sigma_{\rm BAO} = \text{diag}\big(\{\sigma_{{\rm BAO},j}\}\big) \, R_{\rm BAO} \, \text{diag}\big(\{\sigma_{{\rm BAO},j}\}\big) \,,
\end{align}
where $R_{\rm BAO}$ is the correlation matrix relative to the posterior on $\{\theta_{0,j}\}$, see Eq.~\eqref{chi2E}.

\subsection{MCMC exploration}

Summarizing, our analysis consists of two steps.%
\footnote{In Appendix~\ref{ap:alt} we discuss a method with only one step.}
First, we estimate the constraints on $\{ \theta_{{\rm mc},ij} \}$ via Eq.~\eqref{chi2-mes-bao} for all the redshift slices and bin widths. As we will see, this amounts to a total of $17\times7=119$ MCMC evaluations.
Second, we explore the posterior relative to Eq.~\eqref{chi2E} in order to obtain the constraints on $\{\theta_{0,j}\}$ and $E_0$.

For the posterior exploration we use the numerical code \texttt{emcee} \citep{Foreman-Mackey:2012any} and for the visualization of the chains the code \texttt{getdist} \citep{Lewis:2019xzd}.
The convergence of the chains is tested following the  method by \citet{2010CAMCS...5...65G}, which consists in estimating the effective number of independent samples  using the integrated autocorrelation time (which is the mean number of steps needed before the chain forgets where it started).
In the Supplementary Materials we provide plots and convergence tests for all the redshift slices.
In the MCMC exploration, we start the chain from the best-fit model in order to avoid the burn-in phase.

\section{BAO measurements} \label{sec:bao}

\begin{table}
\begin{center}
\setlength{\tabcolsep}{10pt}
\renewcommand{\arraystretch}{1.5}
\begin{tabular}{ccccc}
\hline
$z$ & $\theta_{\rm BAO}$ & $\sigma_{\rm stat}$ & $\sigma_{\rm sys}$ & $\sigma_{\rm BAO}$            \\ 
\hline
0.35 & 5.80 & 0.063 & 0.085 & 0.106 \\
0.37 & 6.07 & 0.103 & 0.089 & 0.136 \\
0.39 & 5.89 & 0.067 & 0.086 & 0.109 \\
0.41 & 5.30 & 0.137 & 0.078 & 0.157 \\
0.43 & 4.87 & 0.060 & 0.071 & 0.093 \\
0.45 & 4.52 & 0.134 & 0.066 & 0.150 \\
0.47 & 4.69 & 0.111 & 0.069 & 0.131 \\
0.49 & 4.69 & 0.041 & 0.069 & 0.080 \\
0.51 & 4.65 & 0.088 & 0.068 & 0.112 \\
0.53 & 4.03 & 0.067 & 0.059 & 0.089 \\
0.55 & 3.56 & 0.058 & 0.052 & 0.078 \\
0.57 & 4.36 & 0.081 & 0.064 & 0.103 \\
0.61 & 3.78 & 0.056 & 0.055 & 0.079 \\
0.63 & 3.90 & 0.057 & 0.057 & 0.080 \\
\hline
\end{tabular}
\end{center}
\caption{This table summarizes our final BAO measurements from the luminous red galaxies of BOSS DR12 and eBOSS DR16. The analysis that led to these constraints did not assume a fiducial cosmological model, except for the part relative to the galaxy mocks. The errors are approximately 3\%. The corresponding correlation matrix $R_{\rm BAO}$ is given in Table~\ref{tab:corr}.}
\label{table:table_final}
\end{table}

\begin{table}
\tiny
\begin{center}
\setlength{\tabcolsep}{2pt}
\renewcommand{\arraystretch}{1.5}
\begin{tabular}{l|llllllllllllll}
\hline
& $\theta_{35}$ & $\theta_{37}$  & $\theta_{39}$  & $\theta_{41}$  & $\theta_{43}$  & $\theta_{45}$  & $\theta_{47}$  & $\theta_{49}$  & $\theta_{51}$  & $\theta_{53}$  & $\theta_{55}$  & $\theta_{57}$  & $\theta_{61}$  &  $\theta_{63}$ \\
\hline
$\theta_{35}$&1 & 0.19 & 0.35 & 0.12 & 0.09 & 0.08 & 0.05 & 0.21 & 0.05 & 0.06 & 0.06 & 0.06 & 0.33 & 0.07 \\
$\theta_{37}$&      & 1 & 0.26 & 0.1 & 0.07 & 0.07 & 0.04 & 0.16 & 0.03 & 0.05 & 0.05 & 0.05 & 0.26 & 0.05 \\
$\theta_{39}$&	  &     & 1 & 0.17 & 0.13 & 0.11 & 0.07 & 0.28 & 0.07 & 0.08 & 0.08 & 0.08 & 0.46 & 0.09 \\
$\theta_{41}$&      &     &     & 1 & 0.04 & 0.04 & 0.02 & 0.1 & 0.02 & 0.03 & 0.03 & 0.03 & 0.17 & 0.03 \\
$\theta_{43}$&	  &     &     &     & 1 & 0.03 & 0.03 & 0.08 & 0.01 & 0.02 & 0.02 & 0.02 & 0.12 & 0.03 \\
$\theta_{45}$&	  &     &     &     &     & 1 & 0.02 & 0.06 & 0.01 & 0.02 & 0.02 & 0.01 & 0.11 & 0.02 \\
$\theta_{47}$&	  &     &     &     &     &     & 1 & 0.04 & 0.01 & 0.02 & 0.01 & 0.02 & 0.06 & 0.02 \\
$\theta_{49}$&	  &     &     &     &     &     &     & 1 & 0.04 & 0.05 & 0.05 & 0.05 & 0.27 & 0.06 \\
$\theta_{51}$&	  &     &     &     &     &     &     &     & 1 & 0.02 & 0.01 & 0.01 & 0.06 & 0.01 \\
$\theta_{53}$&	  &     &     &     &     &     &     &     &     & 1 & 0.02 & 0.01 & 0.08 & 0.01 \\
$\theta_{55}$&	  &     &     &     &     &     &     &     &     &     & 1 & 0.02 & 0.07 & 0.01 \\
$\theta_{57}$&	  &     &     &     &     &     &     &     &     &     &     & 1 & 0.08 & 0.02 \\
$\theta_{61}$&	  &     &     &     &     &     &     &     &     &     &     &     & 1 & 0.09 \\
$\theta_{63}$&	  &     &     &     &     &     &     &     &     &     &     &     &     & 1 \\

\hline
\end{tabular}
\end{center}
\caption{Correlation matrix $R_{\rm BAO}$ relative to the 14 BAO measurements of Table~\ref{table:table_final}. The entry $\theta_{35}$ refers to the $\theta_{\rm BAO}$ estimation at $z=0.35$ and similarly for the other redshifts.}
\label{tab:corr}
\end{table}

\begin{figure}
\centering 
\includegraphics[trim={0 0 0 0}, clip, width=1 \columnwidth]{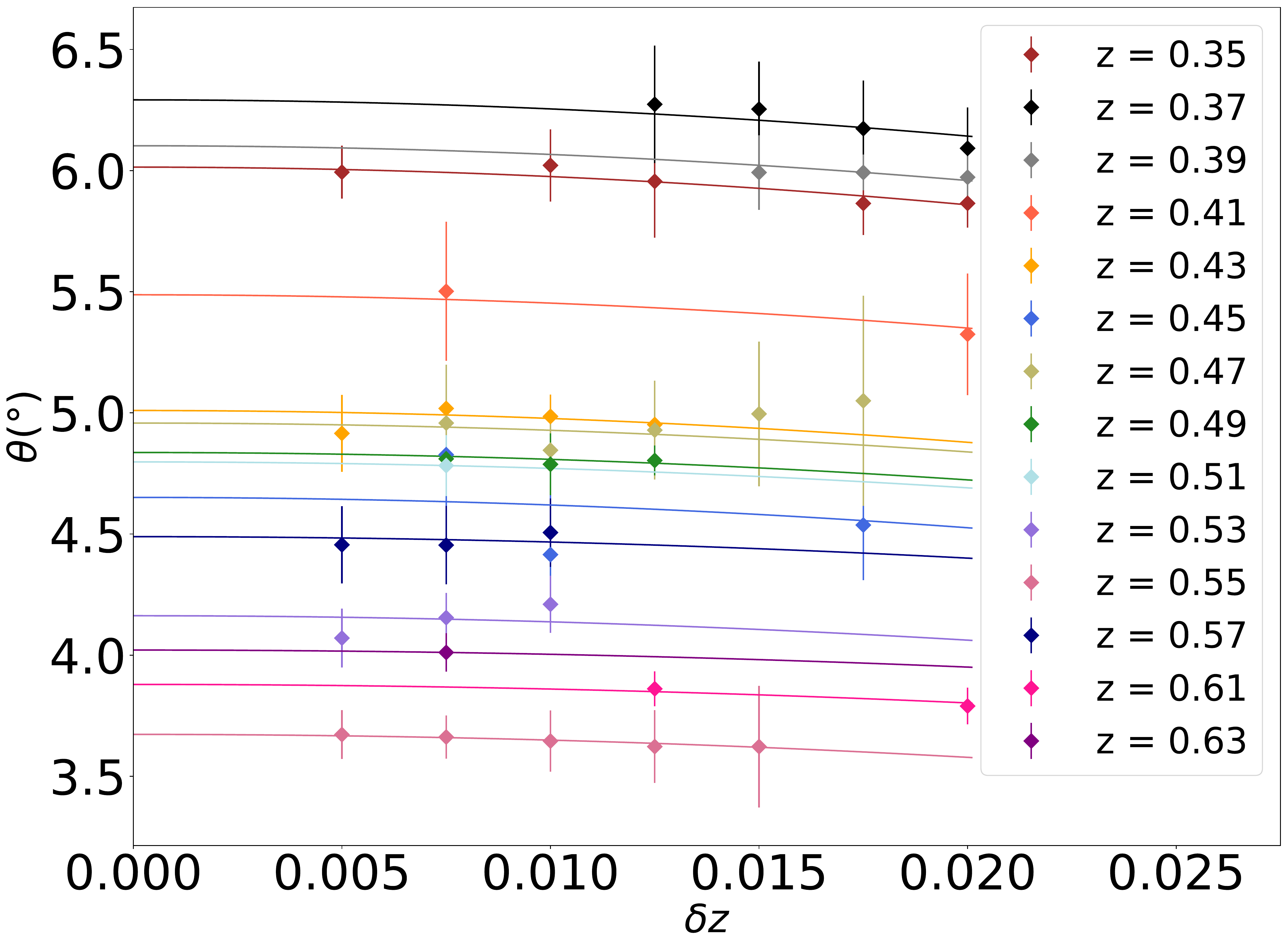}
\caption{
Fit of the model of Eq.~\eqref{eq:Ez} via Eq.~\eqref{chi2E}. Only the measurements with $S_{\rm BAO}>0.95$ are considered, see Eq.~\eqref{S-BAO}.}
\label{fig:Ez}
\end{figure}

We consider the redshift range $0.32<z<0.66$ that has the highest density of galaxies, see Fig.~\ref{fig:boss}.
As we use bins with widths up to $\delta z = 0.02$, we consider the non-overlapping bin centers $0.33, 0.37, \dots, 0.65$.
As said earlier, we only consider measurements with $S_{\rm BAO}>0.95$. None of the measurements relative to the shells at $z=0.33, 0.59, 0.65$ pass this quality cut and are thus absent from the following results, leaving 14 BAO measurements.

Figure~\ref{fig:Ez} shows the best-fit de-projection model for the redshift slices and bin widths that satisfy  $S_{\rm BAO}>0.95$. We find $E_0 = -281 \pm 138$.
Our final BAO results, corrected for projection and modeling biases via Eq.~\eqref{final_bao_report}, with total uncertainties from Eq.~\eqref{total_errors} are given in Table~\ref{table:table_final}.
The correlation matrix $R_{\rm BAO}$ is given in Table~\ref{tab:corr}. The covariance matrix is built via Eq.~\eqref{covabao}.

\begin{figure}
\centering 
\includegraphics[trim={0 0 0 0}, clip, width=1 \columnwidth]{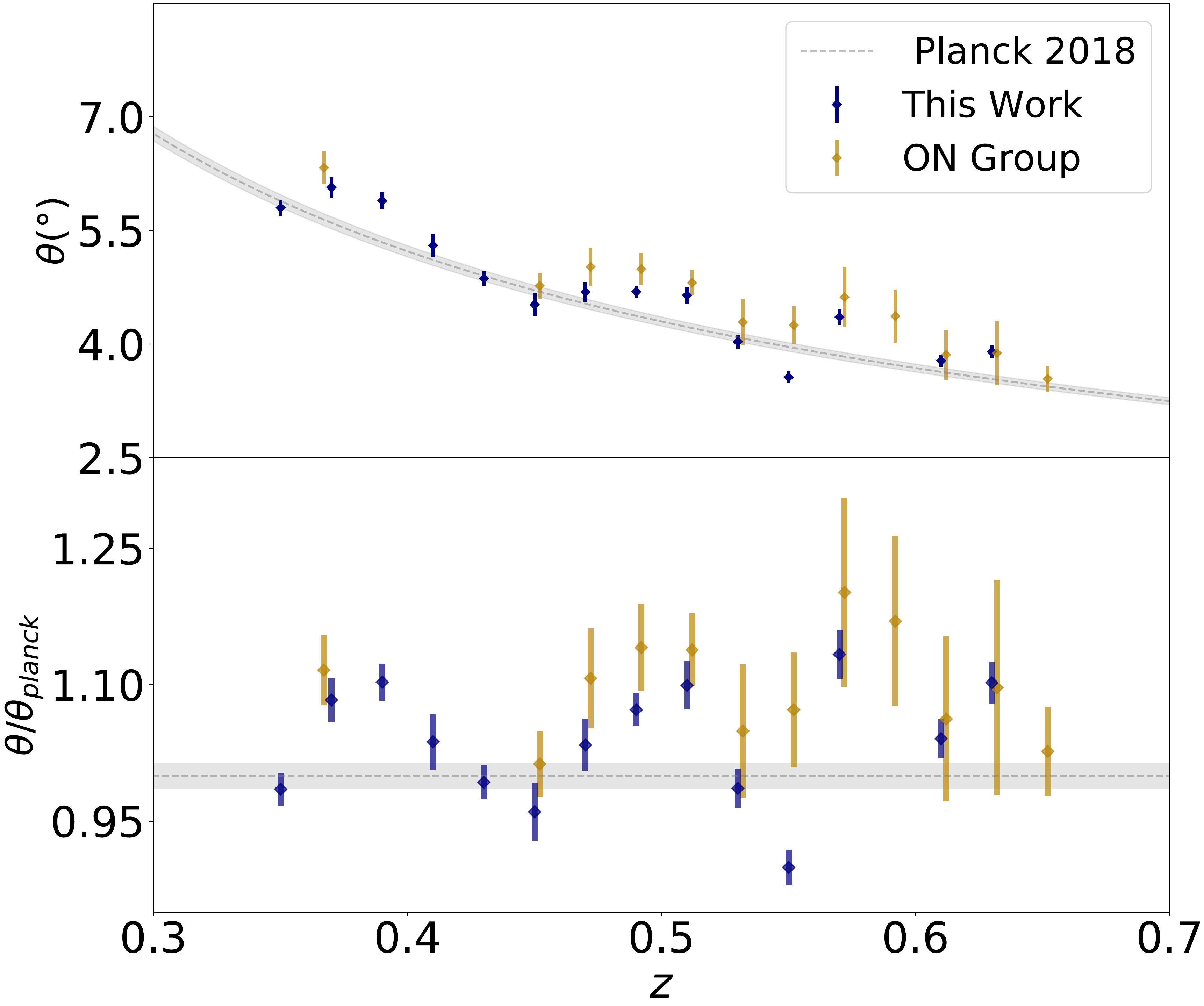}
\caption{
Top panel: Redshift evolution of the BAO measurements of Table~\ref{table:table_final} as compared with the evolution as predicted by the $\Lambda$CDM model with the parameters estimated by Planck 2018~\citep[][Table 2,  last column]{Aghanim:2018eyx}. The gray band corresponds to the $1\sigma$ uncertainty on $r_d \, h$.
Bottom panel: as above, but normalized with respect to the $\Lambda$CDM/Planck 2018 expectation.
}
\label{peaks}
\end{figure}

The evolution of our measurements with the redshift is shown in Fig.~\ref{peaks}.
Also shown are the results that were obtained by the Observatório Nacional (ON) Group \citep{Carvalho:2015ica,Alcaniz:2016ryy,Carvalho:2017tuu}, which used a similar approach to measure the angular BAO scale.
The ON Group found a tension with the results from the CMB; from Fig.~\ref{peaks} it seems that our determinations are closer to the CMB expectation. We will discuss this in more details in the next Section.

\section{Cosmological analysis} \label{sec:cosmo}

We  now use our new BAO measurements of Table~\ref{table:table_final} to constrain the flat $\Lambda$CDM model, whose prediction for $\theta_{\rm BAO}$ is given by Eq.~\eqref{cosmo_BAO}.
The chi-squared function is:
\begin{align}
\chi^2_{\rm BAO}&(H_0, \Omega_m, r_d) \\
&= \sum_{ij} [\theta_{\rm BAO}(z_i) - \theta_{{\rm BAO},i}] \Sigma^{-1}_{{\rm BAO},ij} [\theta_{\rm BAO}(z_j) - \theta_{{\rm BAO},j}] . \nonumber
\label{chi2_bao}
\end{align}

In order to obtain competitive constraints it is useful to include Type Ia Supernovae (SNe) in the analysis. We adopt the Pantheon dataset, consisting of 1048  supernovae spanning the redshift range $0.01 <z< 2.3$~\citep{Scolnic:2017caz}.
The corresponding chi-squared function is:
\begin{eqnarray}
\chi^2_{\rm sne} (H_0, \Omega_m, M) \!=\!\!  \sum_{ij} [ m_{i} \!-\!  m(z_i) ]  S^{-1}_{\text{sne},ij}  [ m_{j} \!-\! m(z_j) ],
\label{chi2_sn}
\end{eqnarray}
where the apparent magnitudes $m_{i}$, redshifts $z_i$ and covariance matrix $S_{\rm sne}$ are from the Pantheon catalog (considering both statistical and systematic errors).

Finally, we will also consider the local prior on the SNe absolute magnitude $M$ relative to the $H_0$ determination by \citet{Riess:2020fzl}, as derived by \citet{Camarena:2019moy,Camarena:2021jlr}:
\begin{equation}
M_{\rm R21}= -19.244 \pm 0.037 \text{ mag} \,.
\label{M_prior}
\end{equation}
The chi-squared function is:
\begin{align}
\chi^2_{\rm R21} (M) = \frac{( M- M_{\rm R21} )^2}{\sigma^2_M} \,. 
\end{align}

\subsection{Angular BAO alone}

\begin{figure}
\centering 
\includegraphics[trim={0 0 0 0}, clip, width=1 \columnwidth]{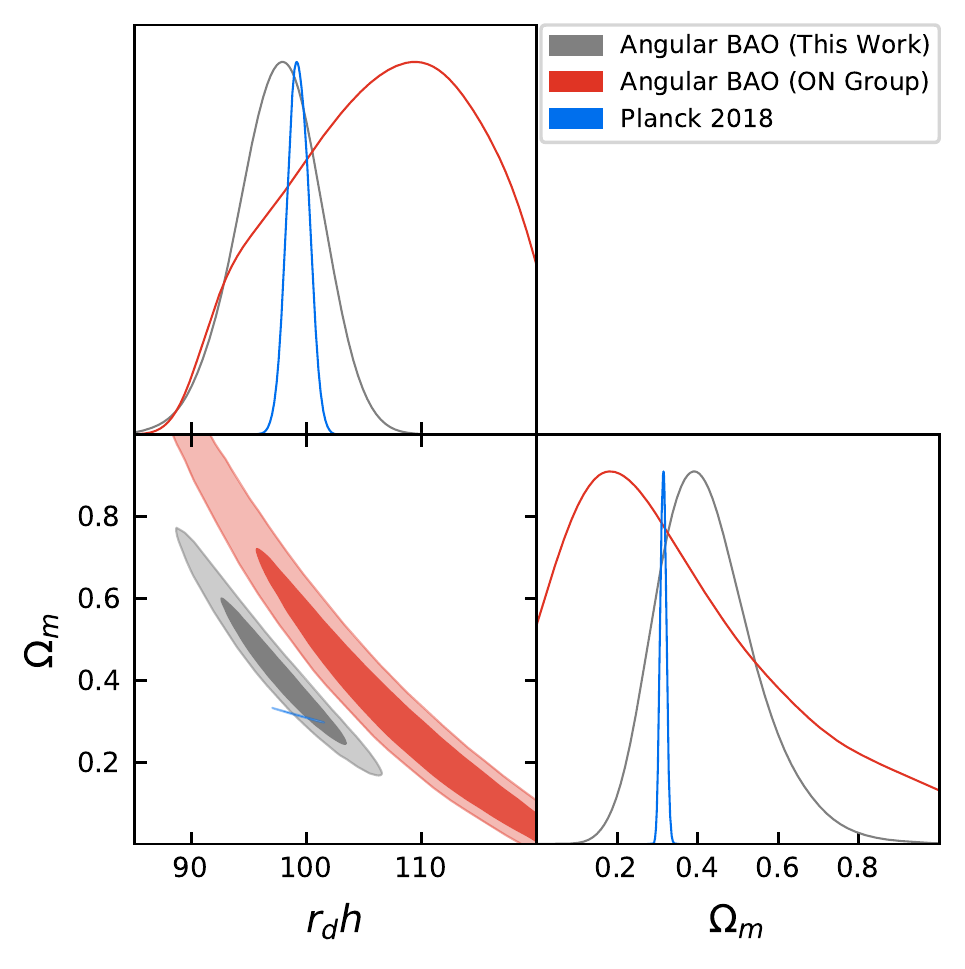}
\caption{Marginalized constraints (68\% and 95\% credible regions) from the BAO measurements of Table~\ref{table:table_final} for the flat $\Lambda$CDM model using the chi-squared function of Eq.~\eqref{BAO-only}.
Also shown are the constraints from the ON Group \citep{Carvalho:2017tuu,Carvalho:2015ica,Alcaniz:2016ryy} and  Planck 2018~\citep[][]{Aghanim:2018eyx}.
}
\label{res:BAO}
\end{figure}

First, we only consider the angular BAO estimates.
Alone, they can only constrain the combination $r_d h$, where $H_0=h \, H_{100}$ and $H_{100} = 100\frac{\rm Km/s}{\rm Mpc}$. Therefore, we adopt the following chi-squared function:
\begin{align} \label{BAO-only}
\tilde \chi^2_{\rm BAO}(r_d h, \Omega_m) = \chi^2_{\rm BAO}(H_{100}, \Omega_m, r_d h ) \,.
\end{align}
Figure~\ref{res:BAO} shows the constraints from our dataset, the ON Group, and  Planck 2018~\citep[][Table 2,  last column]{Aghanim:2018eyx}.
Our constraints are well compatible with the ones from Planck 2018, but in tension with the ones from the ON Group.

\subsection{Angular BAO + SNe}

\begin{figure}
\centering 
\includegraphics[trim={0 0 0 0}, clip, width=1 \columnwidth]{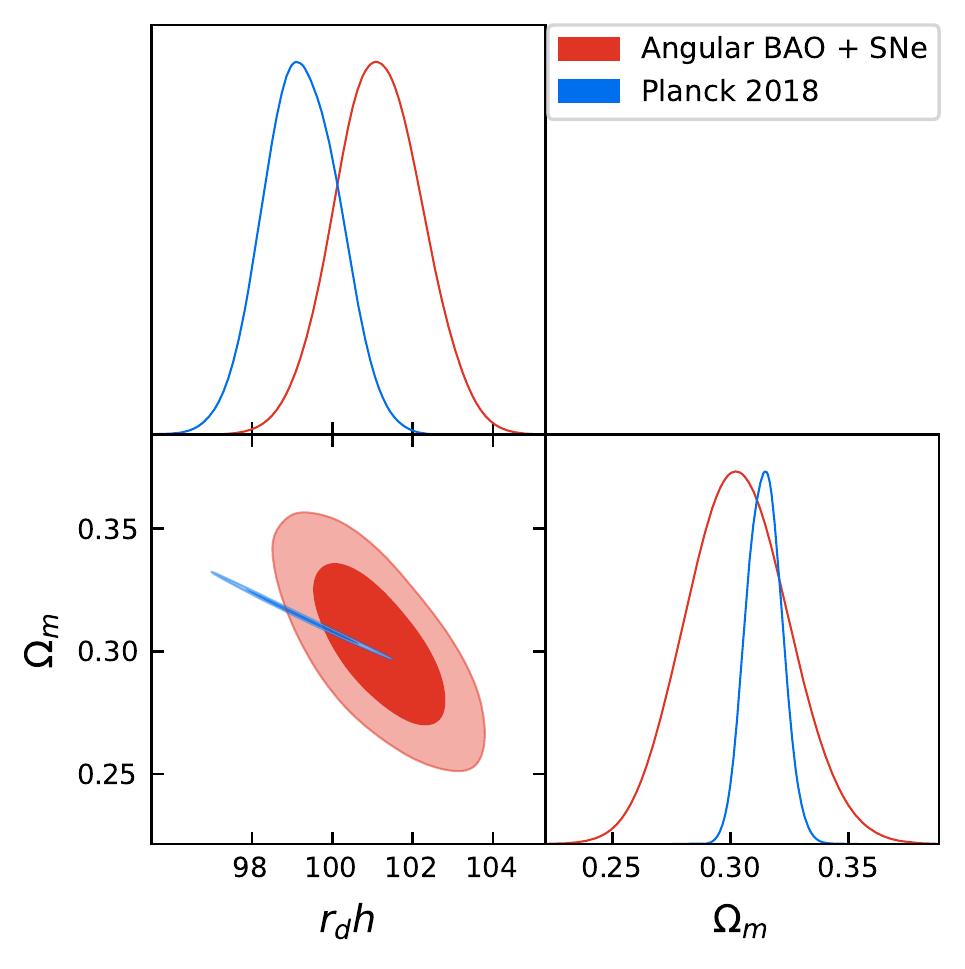}
\caption{Marginalized constraints (68\% and 95\% credible regions) from the BAO measurements of Table~\ref{table:table_final} and the Pantheon supernovae \citep{Scolnic:2017caz} for the flat $\Lambda$CDM model using the chi-squared function of Eq.~\eqref{aBAO-sne}.
Also shown are the constraints from Planck 2018~\citep[][]{Aghanim:2018eyx}.}
\label{flat}
\end{figure}

Next, we include SN data.
Without the use of a prior on $M$ (see next Section), the parameters $H_0$ and $M$ are degenerate. Therefore, it is convenient to adopt the variable $\hat{M} \equiv M -5\log_{10} h$ so that we adopt the following chi-squared function:
\begin{align}
\tilde \chi^2_{\rm sne}(\Omega_m, \hat{M}) = \chi^2_{\rm sne} (H_{100}, \Omega_m, \hat{M}) \,.
\end{align}
The total chi-squared function will be then:
\begin{align} \label{aBAO-sne}
\chi^2(r_d h, \Omega_m, \hat{M}) = \tilde \chi^2_{\rm BAO}(r_d h, \Omega_m) + \tilde \chi^2_{\rm sne}(\Omega_m, \hat{M}) .
\end{align}
Figure~\ref{flat} shows the corresponding constraints together with  the ones from  Planck 2018.
Also in this case we find good agreement.

\subsection{Angular BAO + SNe + SH0ES}
\label{sec:sh0es}

\begin{figure}
\centering 
\includegraphics[trim={0 0 0 0}, clip, width=1 \columnwidth]{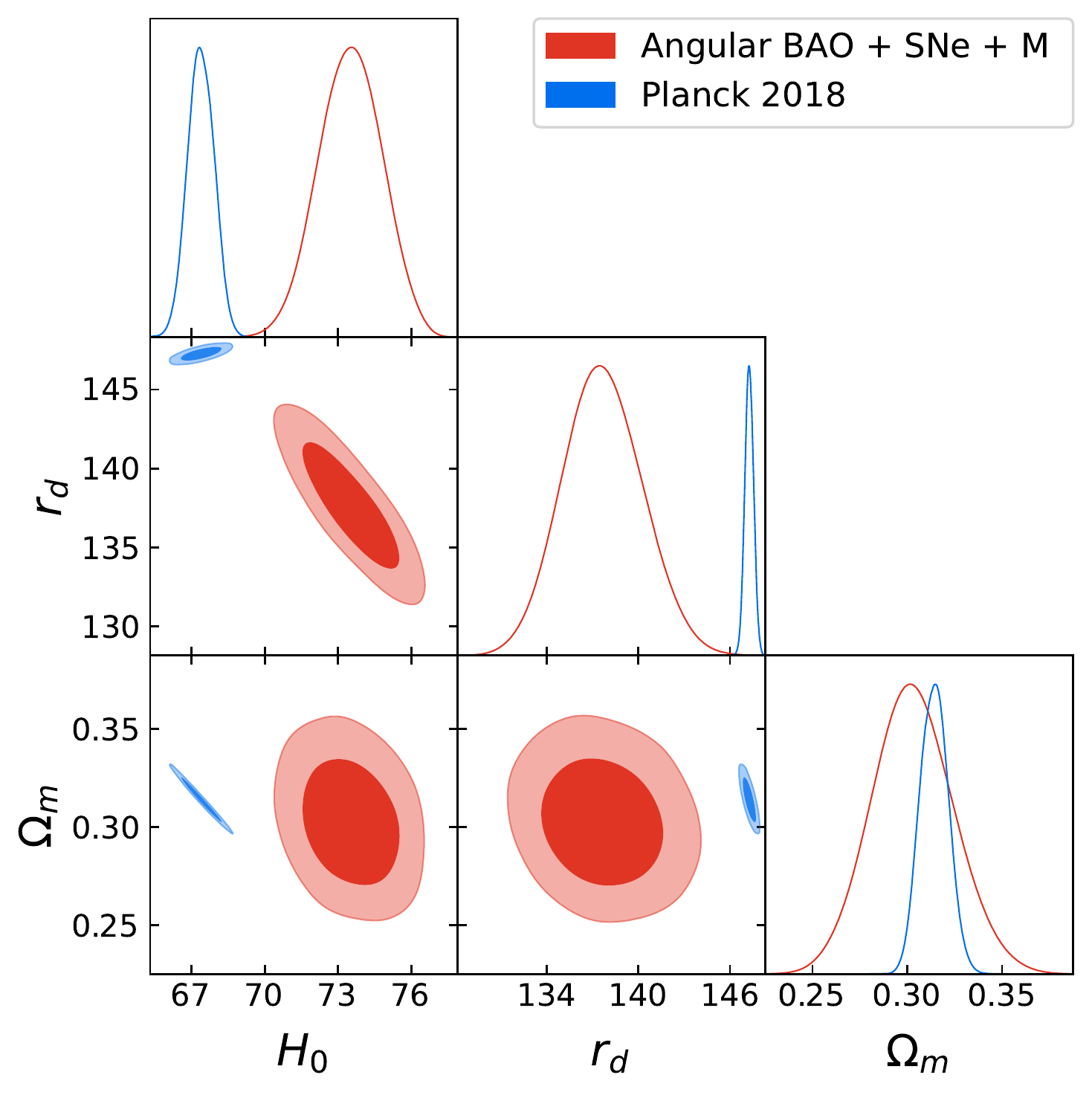}
\caption{
Marginalized constraints (68\% and 95\% credible regions) from the BAO measurements of Table~\ref{table:table_final}, the Pantheon supernovae and the local calibration of the supernova absolute magnitude for the flat $\Lambda$CDM model using the chi-squared function of Eq.~\eqref{chi2all}.
Also shown are the constraints from Planck 2018.
}
\label{SH0ES}
\end{figure}

Finally, we include the local prior on $M$ of Eq.~\eqref{M_prior}. This breaks the degeneracy between $M$ and $H_0$, and this breaks the degeneracy between $H_0$ and $r_d$. The total chi-squared function is then:
\begin{align} \label{chi2all}
\chi^2(h, r_d, \Omega_m, M) = &  \chi^2_{\rm BAO}(H_0, \Omega_m, r_d)   \\
&+  \chi^2_{\rm sne}(H_0, \Omega_m, M)+ \chi^2_{\rm R21} (M) \,. \nonumber
\end{align}
Figure~\ref{SH0ES} shows the corresponding constraints together with  the ones from  Planck 2018.
We see that the SH0ES prior on $M$ causes a strong tension in the $r_d$-$H_0$ plane with respect to Planck 2018: it is the $H_0$ crisis.

\subsection{Tension between angular BAO and CMB}

We show in Table~\ref{table_tension} the tension between the constraints from our angular BAO measurements and the ones from Planck 2018~\citep[][Table 2,  last column]{Aghanim:2018eyx}.
In order to quantity the tension in the $r_d$-$H_0$ plane, we adopt the index of inconsistency (IOI) \citep{Lin:2017ikq}:
\begin{align}
\sqrt{2 \text{IOI}} &\equiv  \sqrt{\delta^T {(C_{\rm BAO} + C_{\rm P18})}^{-1} \delta\,}  \,, \label{tension} \\
\delta &= \{r_d^{\rm BAO}-r_d^{\rm P18},\;  H_0^{\rm BAO} -H_0^{\rm P18} \} \,, \nonumber
\end{align}
where $C$ are the covariance matrices on $r_d$ and $H_0$ from the analysis of Figure~\ref{SH0ES} and $\delta$ is the difference vector.
Note that this estimator assumes Gaussianity and that the posteriors on $r_d$ and $H_0$ are very close to Gaussian. 

We find a good agreement between the $r_d h$ determination from Planck and the ones using our BAO measurements, alone or with supernova data. When we add the local prior on the supernova magnitude we find instead a strong disagreement at the $4.5\sigma$ level.

\begin{table}
\begin{center}
\setlength{\tabcolsep}{3pt}
\renewcommand{\arraystretch}{1.2}
\begin{tabular}{lccc}
\hline
Analysis  & $r_dh $ [Mpc]    & & Tension \\
\hline
Planck 2018   & $99.23 \pm 0.94$ && - \\ 
Angular BAO   & $97.5 \pm 3.6$& & $0.5\sigma$ \\ 
Angular BAO+SNe   & $101.2 \pm 1.1$ && $1.3\sigma$ \\ 
\hline
  & $r_d$ [Mpc] & $H_0 \left[\frac{\rm Km/s}{\rm Mpc}\right]$      & Tension \\
\hline
Planck 2018   & $147.21 \pm 0.23$  &  $67.66  \pm 0.42$  & - \\ 
Angular BAO+SNe+M   & $137.6 \pm 2.6$ & $73.5 \pm 1.3 $ & $4.5\sigma$ \\
\hline
\end{tabular}
\end{center}
\caption{Constraints on $r_d h$, $r_d$ and $H_0$ for the analyses carried out in this paper. The tension in the $r_d$-$H_0$ plane is estimated via the index of inconsistency of Eq.~\eqref{tension}.}
\label{table_tension}
\end{table}

\section{Conclusions}\label{sec:conclusions}

Motivated by the standard model tensions \citep[for a recent review, see][]{Perivolaropoulos:2021jda}, we analyzed BOSS DR12 and eBOSS DR16 galaxies in thin redshift shells, obtaining  a set of 14 BAO measurements with $3\%$ precision.
These results are weakly model dependent as the fiducial model is only introduced when considering the mock catalogs, necessary for the covariance matrix estimation and the pipeline validation.

We find good compatibility with Planck 2018, also when combining these angular BAO measurements with supernova data.
Once we add the SH0ES prior on $M$ we find a strong $4.5\sigma$ tension in the $r_d$-$H_0$ plane with respect to Planck 2018. This model-independent analysis further highlights the $H_0$ crisis.
A Monte Python module for this likelihood is available at \href{https://github.com/ranier137/angularBAO}{github.com/ranier137/angularBAO}.

BAO measurements in thin redshift shells were also obtained by the Observatório Nacional Group, which, however, found a tension with respect to CMB results. This raised questions regarding the consistency of CMB and BAO measurements. According to our results, which are relative to the latest SDSS catalogs, CMB and BAO constraints are compatible.

\section*{Acknowledgements}

It is a pleasure to thank Armando Bernui and Gabriela C.~Carvalho for useful discussions and comments.
RM thanks FAPES for financial support.
VM thanks CNPq and FAPES for partial financial support.
This project has received funding from the European Union’s Horizon 2020 research and innovation programme under the Marie Skłodowska-Curie grant agreement No 888258.
This work also made use of the Virgo Cluster at Cosmo-ufes/UFES, which is funded by FAPES and administrated by Renan Alves de Oliveira.

Funding for SDSS-III has been provided by the Alfred P.\ Sloan Foundation, the Participating Institutions, the National Science Foundation, and the U.S.\ Department of Energy Office of Science. The SDSS-III web site is \href{http://www.sdss3.org/}{sdss3.org}. SDSS-III is managed by the Astrophysical Research Consortium for the Participating Institutions of the SDSS-III Collaboration including the University of Arizona, the Brazilian Participation Group, Brookhaven National Laboratory, Carnegie Mellon University, University of Florida, the French Participation Group, the German Participation Group, Harvard University, the Instituto de Astrofisica de Canarias, the Michigan State/Notre Dame/JINA Participation Group, Johns Hopkins University, Lawrence Berkeley National Laboratory, Max Planck Institute for Astrophysics, Max Planck Institute for Extraterrestrial Physics, New Mexico State University, New York University, Ohio State University, Pennsylvania State University, University of Portsmouth, Princeton University, the Spanish Participation Group, University of Tokyo, University of Utah, Vanderbilt University, University of Virginia, University of Washington, and Yale University.

Funding for the Sloan Digital Sky Survey IV has been provided by the Alfred P.\ Sloan Foundation, the U.S.\ Department of Energy Office of Science, and the Participating Institutions. SDSS-IV acknowledges support and resources from the Center for High Performance Computing at the University of Utah. The SDSS website is \href{www.sdss.org}{sdss.org}. SDSS-IV is managed by the Astrophysical Research Consortium for the Participating Institutions of the SDSS Collaboration including the Brazilian Participation Group, the Carnegie Institution for Science, Carnegie Mellon University, Center for Astrophysics | Harvard \& Smithsonian, the Chilean Participation Group, the French Participation Group, Instituto de Astrof\'isica de Canarias, The Johns Hopkins University, Kavli Institute for the Physics and Mathematics of the Universe (IPMU) / University of Tokyo, the Korean Participation Group, Lawrence Berkeley National Laboratory, Leibniz Institut f\"ur Astrophysik Potsdam (AIP), Max-Planck-Institut f\"ur Astronomie (MPIA Heidelberg), Max-Planck-Institut f\"ur Astrophysik (MPA Garching), Max-Planck-Institut f\"ur Extraterrestrische Physik (MPE), National Astronomical Observatories of China, New Mexico State University, New York University, University of Notre Dame, Observat\'ario Nacional / MCTI, The Ohio State University, Pennsylvania State University, Shanghai Astronomical Observatory, United Kingdom Participation Group, Universidad Nacional Aut\'onoma de M\'exico, University of Arizona, University of Colorado Boulder, University of Oxford, University of Portsmouth, University of Utah, University of Virginia, University of Washington, University of Wisconsin, Vanderbilt University, and Yale University.

The massive production of all MultiDark-Patchy mocks for the BOSS Final Data Release has been performed at the BSC Marenostrum supercomputer, the Hydra cluster at the Instituto de Fısica Teorica UAM/CSIC, and NERSC at the Lawrence Berkeley National Laboratory. We acknowledge support from the Spanish MICINNs Consolider-Ingenio 2010 Programme under grant MultiDark CSD2009-00064, MINECO Centro de Excelencia Severo Ochoa Programme under grant SEV- 2012-0249, and grant AYA2014-60641-C2-1-P. The MultiDark-Patchy mocks was an effort led from the IFT UAM-CSIC by F.\ Prada’s group (C.-H.\ Chuang, S.\ Rodriguez-Torres and C.\ Scoccola) in collaboration with C.\ Zhao (Tsinghua U.), F.-S.\ Kitaura (AIP), A.\ Klypin (NMSU), G.\ Yepes (UAM), and the BOSS galaxy clustering working group.


\section*{Data availability}

The data underlying this article will be shared on reasonable request to the corresponding author.

\bibliographystyle{mnrasArxiv}
\bibliography{biblio}

\appendix


\section{Alternative analysis}\label{ap:alt}

For completeness, we present an alternative analysis which is, however, computationally challenging.
Let us consider, for each redshift $z_j$, the vectors $\tilde \theta_{i}$ and $\tilde w_{j,i}$ of $7 \times 19 = 133$ data points, the 7 correlation functions at the various $\delta z_i$, with the corresponding 19 values of $\theta$. Let us compute the $133\times 133$ covariance matrix $Z_j$ from the mocks.
The model be:
\begin{align}
w_{j}(\tilde \theta)  &=
\begin{cases}
\dots & \dots \\
w_{ji}(\theta) & \text{ if } \delta z=\delta z_i \\ 
\dots & \dots
\end{cases} \,, \\
w_{ji}(\theta) & = A_{ji} + B_{ji}\theta^{\nu_{ji}} + C_{ji} e^{-\frac{(\theta-\Theta_{ji})^2}{2\sigma_{ji}^2}} \,, \\
\Theta_{ji}  &= \theta_{0,j} + E(z_j)\delta z_i^2  \,, \\
E(z_j) & = E_0 + E_1 z_j \,.
\label{phenom_model2}
\end{align}
The chi-squared function is:
\begin{align}
\chi^2 &= \sum_{j=1}^{15}  \chi_j^2  \,, \\
\chi_j^2 &= \sum_{i=1,k=1}^{133,133}
\left[\tilde w_{j,i}- w_j(\tilde \theta_{i})\right] Z^{-1}_{j, ik}\left[\tilde w_{j,k}- w_j(\tilde \theta_{k})\right]  \,.
\end{align}

In this way, one constrains the 17 BAO values $\theta_{0,j}$ via one single analysis.
One can then apply quality cuts via $S_{{\rm BAO},ji}$.
The drawback is that it involves $(5 \times 7 \times 17  + 2) = 597$ nuisance parameters and, therefore, a 614-dimensional parameter space.

\bsp	
\label{lastpage}
\end{document}